\documentclass[12pt]{iopart}

\usepackage{graphicx}
\usepackage{bm}
\usepackage[squaren, thinqspace]{SIunits}
\usepackage{epsfig}
\usepackage{xspace}
\usepackage{iopams}
\usepackage{setstack}
\usepackage[T1]{fontenc}

\begin{document}

\renewcommand{\degree}{\ensuremath{^\circ}\xspace}
\newcommand{\Vp}{\ensuremath{V_\mathrm{p}}\xspace}
\newcommand{\Hy}{\ensuremath{\mathbf{H} || \mathbf{\hat{y}}}\xspace}
\newcommand{\Hx}{\ensuremath{\mathbf{H} || \mathbf{\hat{x}}}\xspace}
\newcommand{\y}{\ensuremath{\mathbf{\hat{y}}}\xspace}
\newcommand{\x}{\ensuremath{\mathbf{\hat{x}}}\xspace}
\newcommand{\z}{\ensuremath{\mathbf{\hat{z}}}\xspace}
\newcommand{\FPH}{ferromagnetic thin film/piezoelectric actuator hybrid\xspace}
\newcommand{\eqref}{\eref}

\title[Voltage controlled inversion of magnetic anisotropy]{Voltage controlled inversion of magnetic anisotropy in a ferromagnetic thin film at room temperature}

\author{M Weiler$^1$, A Brandlmaier$^1$, S Gepr\"{a}gs$^1$, M Althammer$^1$, M Opel$^1$, C Bihler$^2$, H Huebl$^{2,4}$, M S Brandt$^2$, R Gross$^{1,3}$ and S T B Goennenwein$^{1,3}$}
\address{$^1$Walther-Mei{\ss}ner-Institut, Bayerische Akademie der Wissenschaften, 85748 Garching, Germany}
\address{$^2$Walter Schottky Institut, Technische Universit\"{a}t M\"{u}nchen, 85748 Garching, Germany}
\address{$^3$Physik-Department, Technische Universit\"{a}t M\"{u}nchen, 85748 Garching, Germany}
\address{$^4$Present address: Australian Research Council Centre of Excellence for Quantum Computer Technology, School of Physics, The University of New South Wales, Sydney, NSW 2052, Australia.}
\ead{mathias.weiler@wmi.badw.de}

\date{\today}

\begin{abstract}
The control of magnetic properties by means of an electric field is an important aspect in magnetism and magnetoelectronics. We here utilize magnetoelastic coupling in ferromagnetic/piezoelectric hybrids to realize a voltage control of magnetization orientation at room temperature. The samples consist of polycrystalline nickel thin films  evaporated onto piezoelectric actuators. The magnetic properties of these multifunctional hybrids are investigated at room temperature as a function of the voltage controlled stress exerted by the actuator on the Ni film.
Ferromagnetic resonance spectroscopy shows that the magnetic easy axis in the Ni film plane is rotated by 90\degree upon changing the polarity of the voltage \Vp applied to the actuator. In other words, the in-plane uniaxial magnetic anisotropy of the Ni film can be inverted via the application of an appropriate voltage \Vp. Using SQUID magnetometry, the evolution of the magnetization vector is recorded as a function of \Vp and of the external magnetic field. Changing \Vp allows to reversibly adjust the magnetization orientation in the Ni film plane within a range of approximately 70\degree. All magnetometry data can be quantitatively understood in terms of the magnetic free energy determined from the ferromagnetic resonance experiments. These results demonstrate that magnetoelastic coupling in hybrid structures indeed is a viable option to control magnetization orientation in technologically relevant ferromagnetic thin films at room temperature.
\end{abstract}

\maketitle
\section{Introduction}
Multifunctional material systems unite different, e.g., electric and magnetic, functionalities in either a single phase or a heterostructure. They thus are of great fundamental and technological interest. Given that the electric and magnetic properties furthermore are coupled, it becomes possible to control either the magnetization by the application of an electric field or the electric polarization via a magnetic field alone. An electric field control of magnetization is particularly appealing, as it removes the need to generate magnetic fields of sufficient strength for magnetization switching on small length scales -- thus enabling novel concepts for high density magnetic data storage applications.
\begin{figure}
\centering
  \includegraphics[]{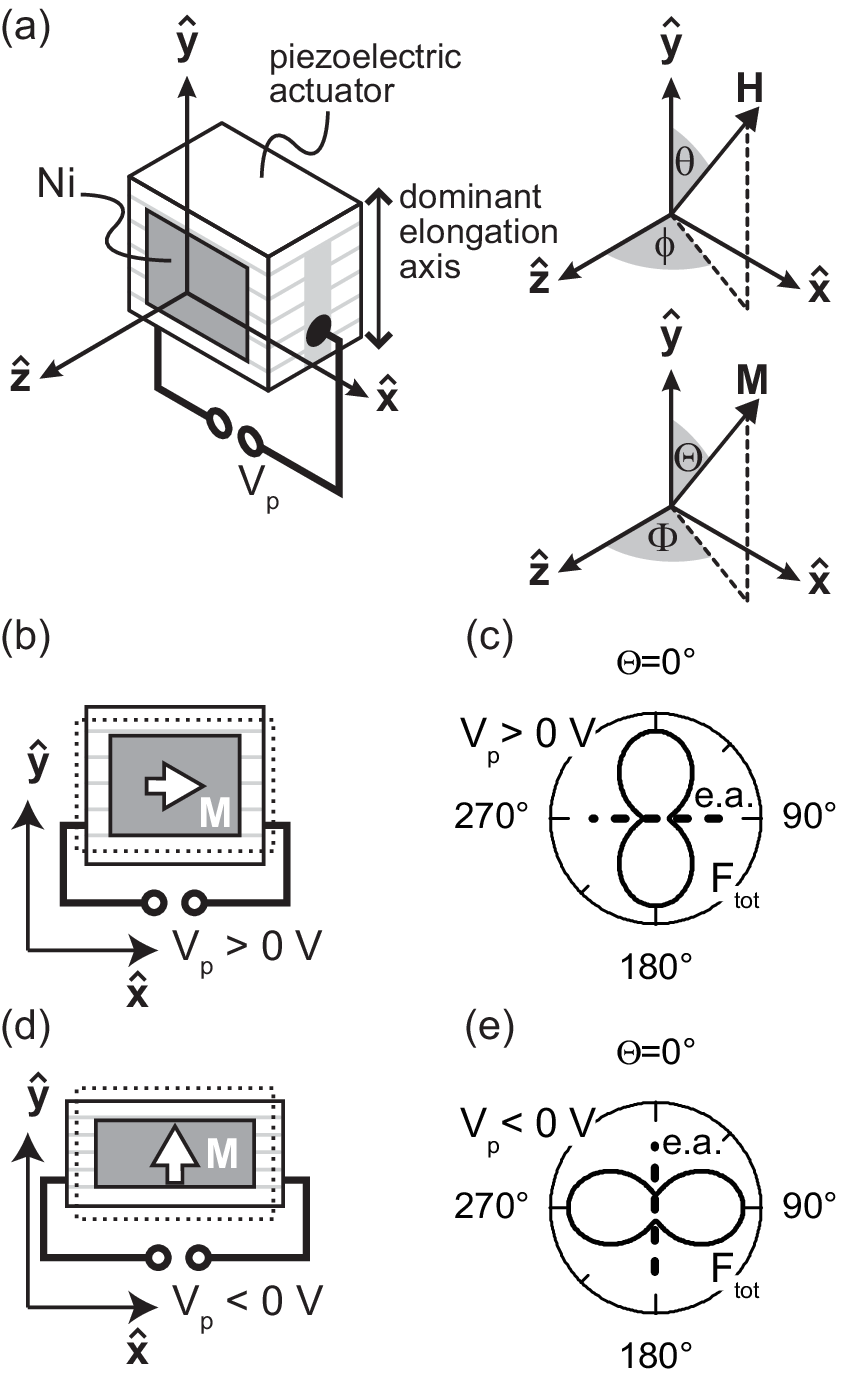}\\
  \caption{(a) Schematic illustration of the \FPH. (b), (d) The application of a voltage $\Vp\neq\unit{0}{\volt}$ to the actuator results in a deformation of the actuator and the affixed ferromagnetic film. The relaxed actuator at $\Vp=\unit{0}{\volt}$ is shown by dotted contours. (c), (e) Schematic free energy contours in the film plane.  The magnetic easy axis (e.a.) shown by the thick dashed line is oriented parallel to the compressive strain and can thus be rotated by 90\degree by changing the polarity of $\Vp$.}\label{fig:panel1}
\end{figure}
Thus, techniques to control the magnetization by means of electric fields or currents have recently been vigorously investigated and several schemes for an electric control of magnetization have been reported. These include the spin-torque effect~\cite{Slonczewski:1996, berger:1996, tsoi:1998, zhang:2002} in spin-valves and magnetic tunnel junctions and the direct electric field control of magnetization in intrinsically multiferroic materials~\cite{spaldin:multiferroics, Eerenstein:2006, Fiebig2005,ramesh:2007,Lottermoser2004, gajek:2007, Chu2008, Zhao:2006} or in ferromagnetic/ferroelectric heterostructures~\cite{Stolichnov:2008, binek:2005}. A third, very attractive approach for the electric field control of magnetization takes advantage of the elastic channel, i.e., magnetostrictive coupling~\cite{bootsmann:2005, wan:2006, Doerr:2006, Goennenwein:2008, brandlmaier:2008, Bihler:2008, overby:2008, jungwirth:2008, wang:2005}.
We here show that magnetoelasticity -- that is the effect of lattice strain on magnetic anisotropy -- in a polycrystalline nickel film/piezoelectric actuator hybrid structure allows to switch the magnetic easy axis in the Ni film by 90\degree at room temperature by simply changing the polarity of the voltage applied to the actuator.
This can be used to achieve either an irreversible magnetization orientation control of 180\degree or a reversible control close to 90\degree, depending on whether the magnetization orientation is prepared in a local or global minimum of the free energy via a magnetic field sweep prior to the electric control experiment.
We use ferromagnetic resonance (FMR) to quantify the effect of an electric field on magnetic anisotropy and superconducting quantum interference device (SQUID) magnetometry to directly record the evolution of magnetization as a function of the applied electric field. These experiments demonstrate that the strain-mediated electric field control of magnetization indeed is a viable technique in technologically relevant ferromagnetic thin films at room temperature.
\section{Sample preparation and experimental techniques}\label{sec:samples}
We fabricated ferromagnetic thin film/piezoelectric actuator structures by depositing nickel (Ni) thin films onto piezoelectric Pb(Zr$_x$Ti$_{1-x}$)O$_3$-based actuators~\cite{manual:piezo}. Nickel was chosen as the ferromagnetic constituent as it is a prototype 3d itinerant ferromagnet with a Curie temperature $T_\mathrm{c}=\unit{627}{\kelvin}$ well above room temperature~\cite{kittel:ssp}, a high bulk saturation magnetization $M_\mathrm{s}=\unit{411}{\kilo\ampere\per\meter}$~\cite{danan:1968} and sizeable volume magnetostriction~\cite{chikazumi:ferromagnetism, lee:1955} $\overline{\lambda}=\frac{2}{5}\lambda_{100}+\frac{3}{5}\lambda_{111}=-32.9\times10^{-6}$ with $\lambda_{100}$ and $\lambda_{111}$ being the single crystal saturation magnetostriction for a magnetic field applied along a crystalline $<100>$ or $<111>$ axis, respectively.
The actuators exhibit a hysteretic mechanical stroke of up to $1.3\times10^{-3}$ along their dominant elongation axis [cf. Fig.~\ref{fig:panel1}(a)] if voltages $\unit{-30}{\volt}\leq \Vp \leq\unit{+150}{\volt}$ are applied.
Prior to the deposition of the Ni film, the actuators were mechanically polished to a size of $x\times y\times z=\unit{3\times2.6\times2}{\milli\meter\cubed}$ [cf. Fig.~\ref{fig:panel1}(a)] to accommodate the size of the sample to the restrictions imposed by the ferromagnetic resonance setup. We then used electron beam evaporation at a base pressure of \unit{2.0\times10^{-8} }{\milli\bbar} to deposit a \unit{70}{\nano\meter} thick Ni film onto an area of \unit{5}{\milli\meter\squared} on the \x-\y face of the  actuators. To prevent oxidation of the Ni film, a \unit{10}{\nano\meter} thick Au film was deposited in-situ on top of the Ni layer. The multifunctional hybrid obtained after Ni deposition is sketched schematically in Fig.~\ref{fig:panel1}(a) together with the definition of the angles that describe the orientation of the magnetization $\mathbf{M}=(M,\Theta,\Phi)$ and the external magnetic field $\mathbf{H}=(H,\theta,\phi)$ in the sample-affixed coordinate system.

To determine the static magnetic response of the \FPH  we employ superconducting quantum interference device (SQUID) magnetometry. The Quantum Design MPMS-XL-7 SQUID magnetometer is sensitive to the projection $m=\mathbf{m}\mathbf{\hat{H}}$ of the total magnetic moment $\mathbf{m}$ onto the unit vector $\mathbf{\hat{H}}=\frac{\mathbf{H}}{H}$. We corrected $m$ for the paramagnetic contribution of the actuator and used the Ni film volume $V=\unit{3.5\times10^{-13}}{\meter\cubed}$ to calculate the respective projection $M=m/V$ of the magnetization onto $\mathbf{\hat{H}}$. All magnetometry data shown in the following were recorded at a temperature $T=\unit{300}{\kelvin}$.

The magnetic anisotropy of the \FPH was measured  by ferromagnetic resonance (FMR) at room temperature. We use a Bruker ESP 300 spin resonance spectrometer with a TE$_{102}$ cavity operating at a constant microwave frequency $\nu_\mathrm{MW}=\unit{9.3}{\giga\hertz}$. The sample can be rotated in the FMR setup with respect to the external dc magnetic field, so that either $\theta$ or $\phi$ [cf. Fig.~\ref{fig:panel1}(a)] can be adjusted at will. To allow for lock-in detection we use magnetic field modulation at a modulation frequency of \unit{100}{\kilo\hertz} with a modulation amplitude of $\mu_0 H_\mathrm{mod}=\unit{3.2}{\milli\tesla}$.

\section{Phenomenology of strain-induced magnetic anisotropy}\label{sec:theory}

The piezoelectric actuator deforms upon the application of a voltage $\Vp \neq \unit{0}{\volt}$. Due to its elasticity, an elongation (contraction) along one cartesian direction is always accompanied by a contraction (elongation) in the two orthogonal directions. Therefore, for $\Vp>\unit{0}{\volt}$, the actuator expands along its dominant elongation axis \y and contracts along the two orthogonal directions \x and \z. The Ni film affixed to the \x-\y face of the actuator is hence strained tensilely along \y and compressively along \x for $\Vp>\unit{0}{\volt}$ [cf. Fig.~\ref{fig:panel1}(b)]. For $\Vp<\unit{0}{\volt}$ the actuator contracts along \y and thus expands along \x and \z  and the Ni film thus exhibits a compressive strain along \y and a tensile strain along \x [cf. Fig.~\ref{fig:panel1}(d)].

To describe the impact of this lattice strain on the Ni magnetization orientation we use a magnetic free energy density approach. The free energy $F_\mathrm{tot}$ is a measure for the angular dependence of the magnetic hardness, with maxima in $F_\mathrm{tot}$ corresponding to magnetically hard directions and minima to magnetically easy directions. In equilibrium, the magnetization always resides in a local minimum of $F_\mathrm{tot}$. Contrary to single-crystalline films, the evaporated Ni films are polycrystalline and thus show no net crystalline magnetic anisotropy which may compete with strain-induced anisotropies and hereby reduce the achievable magnetization orientation effect~\cite{brandlmaier:2008}. $F_\mathrm{tot}$ is thus given by
\begin{equation}\label{eq:Ftot}
  F_\mathrm{tot}=F_\mathrm{stat}+F_\mathrm{demag}+F_\mathrm{magel}\;.
\end{equation}
The first term $F_\mathrm{stat} = -\mu_0 M H(\sin\Theta \sin\Phi \sin\theta \sin\phi +\cos\Theta \cos\theta+\sin\Theta \cos\Phi \sin\theta \cos\phi)$ in Eq.~\eqref{eq:Ftot} is the Zeemann term and  describes the influence of an external magnetic field $\mathbf{H}$ on the orientation of $\mathbf{M}$. The uniaxial demagnetization term $F_\mathrm{demag}=\frac{\mu_0}{2} M^2 \sin^2\Theta\cos^2\Phi$ is the anisotropy caused by the thin-film shape of the sample~\cite{Morrish2001}. The last contribution to Eq.~\eqref{eq:Ftot}
\begin{eqnarray}\label{eq:Fmagel}
    F_\mathrm{magel}&=&\frac{3}{2} \overline{\lambda} \left(c^\mathrm{Ni}_{12}-c^\mathrm{Ni}_{11}\right)
  \bigl[\varepsilon_1(\sin^2\Theta\sin^2\Phi-1/3) \\
   &&+\varepsilon_2(\cos^2\Theta-1/3)+\varepsilon_3(\sin^2\Theta\cos^2\Phi-1/3) \bigr]\nonumber
\end{eqnarray}
describes the influence of the lattice strains on the magnetic anisotropy~\cite{chikazumi:ferromagnetism}. The strains along the \x-, \y- and \z-axis are denoted in Voigt (matrix) notation~\cite{nye:1985} as $\varepsilon_1, \varepsilon_2$ and $\varepsilon_3$, respectively. Furthermore, $c^\mathrm{Ni}_{11}=\unit{2.5\times10^{11}}{\newton\per\meter^2}$ and $c^\mathrm{Ni}_{12}=\unit{1.6\times10^{11}}{\newton\per\meter^2}$ are the elastic moduli of Ni~\cite{lee:1955}. The effects of shear strains ($\varepsilon_i, i \in \{4,5,6\}$) average out in our polycrystalline film and thus are neglected.

Using $\sin^2\Theta\sin^2\Phi+\cos^2\Theta+\sin^2\Theta\cos^2\Phi=1$ and omitting isotropic terms, Eq.~\eqref{eq:Fmagel} can be rewritten as
\begin{equation}\label{eq:Fmagel_final}
    F_\mathrm{magel}=K_\mathrm{u,magel,y} \cos^2\Theta + K_\mathrm{u,magel,z} \sin^2\Theta \cos^2\Phi
\end{equation}
with
\begin{eqnarray}\label{eq:Kmagel}
  K_\mathrm{u,magel,y} &=& \frac{3}{2} \overline{\lambda}(c^\mathrm{Ni}_{12}-c^\mathrm{Ni}_{11})(\varepsilon_2-\varepsilon_1) \\
 \nonumber K_\mathrm{u,magel,z} &=& \frac{3}{2} \overline{\lambda} (c^\mathrm{Ni}_{12}-c^\mathrm{Ni}_{11})(\varepsilon_3-\varepsilon_1)\;.
\end{eqnarray}

Due to the elasticity of the actuator and the Ni film, the strains $\varepsilon_i$ are not independent of each other. The strains $\varepsilon_1$ and $\varepsilon_2$ in the \x-\y plane are linked by the Poisson ratio $\nu=0.45$ of the actuator~\cite{manual:piezo} according to
\begin{equation}\label{eq:poisson_ip}
    \varepsilon_1=-\nu \varepsilon_2 \;.
\end{equation}
Furthermore, due to the elasticity of the Ni film, the out-of-plane strain $\varepsilon_3$ can be expressed as~\cite{brandlmaier:2008}
\begin{equation}\label{eq:strain_oop}
    \varepsilon_3=-\frac{c^\mathrm{Ni}_{12}}{c^\mathrm{Ni}_{11}}(\varepsilon_1+\varepsilon_2) \;.
\end{equation}
Assuming a linear expansion of the actuator with lateral dimension $L$ parallel to its dominant elongation axis, we can calculate the strain in the Ni film parallel to the actuator's dominant elongation axis as
\begin{equation}\label{eq:strain_voltage}
    \varepsilon_2=\frac{\delta L}{L}\frac{\Vp}{\unit{180}{\volt}}\;,
\end{equation}
where $\delta L /L=1.3\times10^{-3}$  is the nominal full actuator stroke for the full voltage swing $\unit{-30}{\volt}\leq \Vp \leq\unit{+150}{\volt}$.
With Eqs.~\eqref{eq:poisson_ip}, \eqref{eq:strain_oop} and~\eqref{eq:strain_voltage} all strains in the Ni film can thus directly be calculated for any given voltage \Vp. This allows to determine the equilibrium orientation of the magnetization as a function of \Vp by minimizing the total magnetoelastic free energy density $F_\mathrm{tot}$ [Eq.~\eqref{eq:Ftot}].

Due to the negative magnetostriction ($\overline{\lambda}<0$) of Ni and $c^\mathrm{Ni}_{11}>c^\mathrm{Ni}_{12}$, the in-plane easy axis of $F_\mathrm{tot}$ is oriented orthogonal to tensile and parallel to compressive strains in the absence of external magnetic fields. Due to the strong uniaxial out-of-plane anisotropy caused by $F_\mathrm{demag}$, the in-plane easy axis is the global easy axis. For $\Vp>\unit{0}{\volt}$,  the Ni film exhibits a tensile strain along \y ($\varepsilon_2>0$) and a compressive strain along \x  ($\varepsilon_1<0$). The easy axis of $F_\mathrm{tot}$ is thus oriented along the \x-direction [cf. Fig.~\ref{fig:panel1}(c)]. Accordingly, for $\Vp<\unit{0}{\volt}$, the easy axis of $F_\mathrm{tot}$ is oriented parallel to the compressive strain along \y and orthogonal to the tensile strain along \x [cf. Fig.~\ref{fig:panel1}(e)]. We thus expect a 90\degree rotation of the in-plane easy axis of $F_\mathrm{tot}$ upon changing the polarity of \Vp.
Note that in the absence of external magnetic fields there are always two energetically equivalent antiparallel orientations of $\mathbf{M}$ along the magnetic easy axis. However, for simplicity, we only show one of the two resulting possible in-plane orientations of $\mathbf{M}$ in Figs.~\ref{fig:panel1}(b) and~\ref{fig:panel1}(d).
Nevertheless these panels show that it should be possible to change the orientation of $\mathbf{M}$ from $\mathbf{M} || \x$ to $\mathbf{M} || \y$ via the application of appropriate voltages $\Vp$ to the actuator.

The total magnetic free energy density $F_\mathrm{tot}$ in Eq.~\eqref{eq:Ftot} is experimentally accessible by ferromagnetic resonance measurements. The FMR equations of motion~\cite{smit:1954, smit:1955, suhl:1955}
\begin{eqnarray}\label{eq:FMR_motion}
	 \left(\frac{\omega}{\gamma}\right)^2&=&\frac{1}{M_\mathrm{s}^2\sin^2\Theta} \bigl.\Bigl(\left(\partial_{\Phi}^2F_\mathrm{tot}\right)\left(\partial_\Theta^2 F_\mathrm{tot}\right)-\left(\partial_{\Phi}\partial_\Theta F_\mathrm{tot}\right)^2\Bigr)\bigr|_{\Theta_0,\Phi_0}\\
\nonumber &&\mathrm{and}~\bigl.\partial_\Theta F_\mathrm{tot}\bigr|_{\Theta=\Theta_0} = \bigl.\partial_{\Phi} F_\mathrm{tot}\bigr|_{\Phi={\Phi}_0}=0
    \end{eqnarray}
link the experimentally determined ferromagnetic resonance field $\mu_0H_\mathrm{res}$ to the magnetic free energy density $F_\mathrm{tot}$. In Eq.~\eqref{eq:FMR_motion}, $\gamma$ is the gyromagnetic ratio and $\omega=2\pi\nu_\mathrm{MW}$ with the microwave frequency $\nu_\mathrm{MW}$.
The effect of damping is neglected in Eq.~\eqref{eq:FMR_motion}, as damping only affects the lineshape which will not be discussed in this article. We note that the FMR resonance field $\mu_0H_\mathrm{res}$  measured in experiment can be considered as a direct indicator of the relative magnetic hardness along the external dc magnetic field direction. From Eqs.~\eqref{eq:Ftot} and~\eqref{eq:FMR_motion} one finds that smaller resonance fields correspond to magnetically easier directions and larger resonance fields to magnetically harder directions.
\section{Ferromagnetic resonance}\label{sec:fmr}
\begin{figure}
  \includegraphics[width=\textwidth]{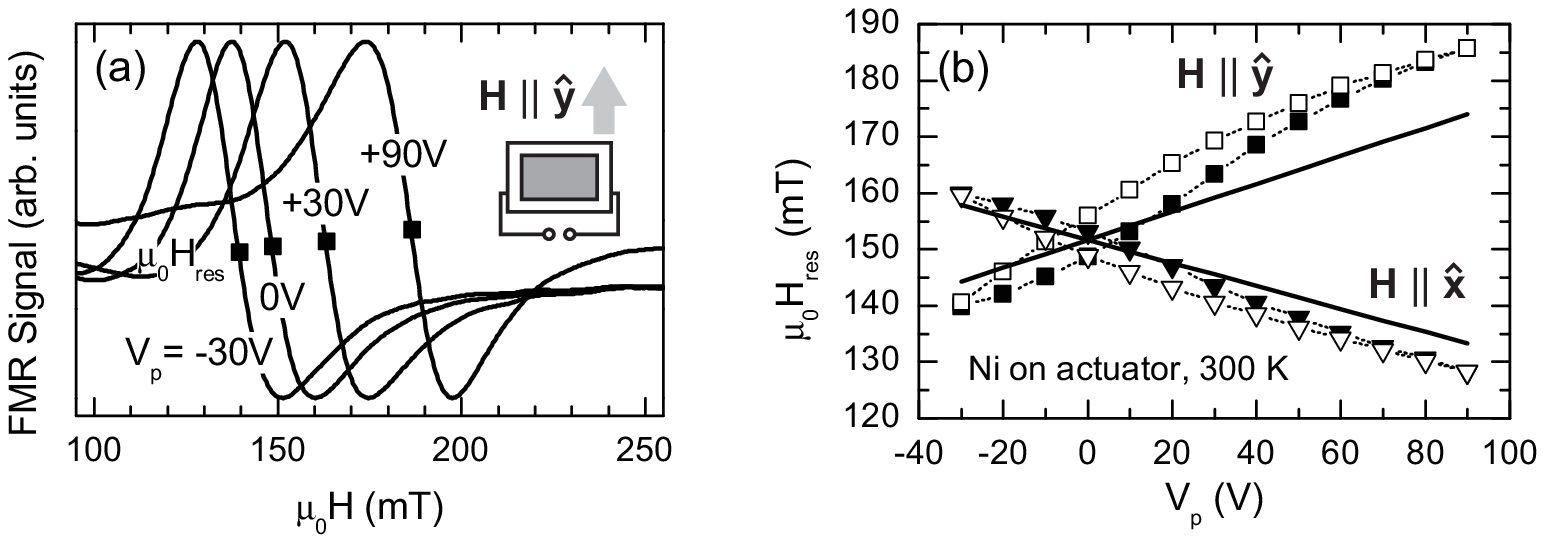}\\
  \caption{(a) FMR spectra recorded with \Hy at different voltages \Vp. An increasing resonance field $\mu_0 H_\mathrm{res}$ (solid squares) is observed for increasing $\Vp$ while the lineshape is not significantly altered. (b) The dependence of $\mu_0 H_\mathrm{res}$ on \Vp is qualitatively different for \Hy and \Hx.  Full symbols correspond to increasing $\Vp$ and open symbols to decreasing $\Vp$. The solid lines represent the resonance fields calculated from magnetoelastic theory (cf. Sec.~\ref{sec:theory}) yielding very good agreement with the measurement. For \Hy, increasing \Vp increases $\mu_0 H_\mathrm{res}$ and thus the magnetic hardness of this direction, while for \Hx, increasing \Vp decreases $\mu_0 H_\mathrm{res}$ and thus the magnetic hardness of this direction. The hysteresis of $\mu_0 H_\mathrm{res}$ is due to the hysteretic stress-strain curve of the actuator.}\label{fig:panel2}
\end{figure}
\begin{figure}
  \includegraphics[width=\textwidth]{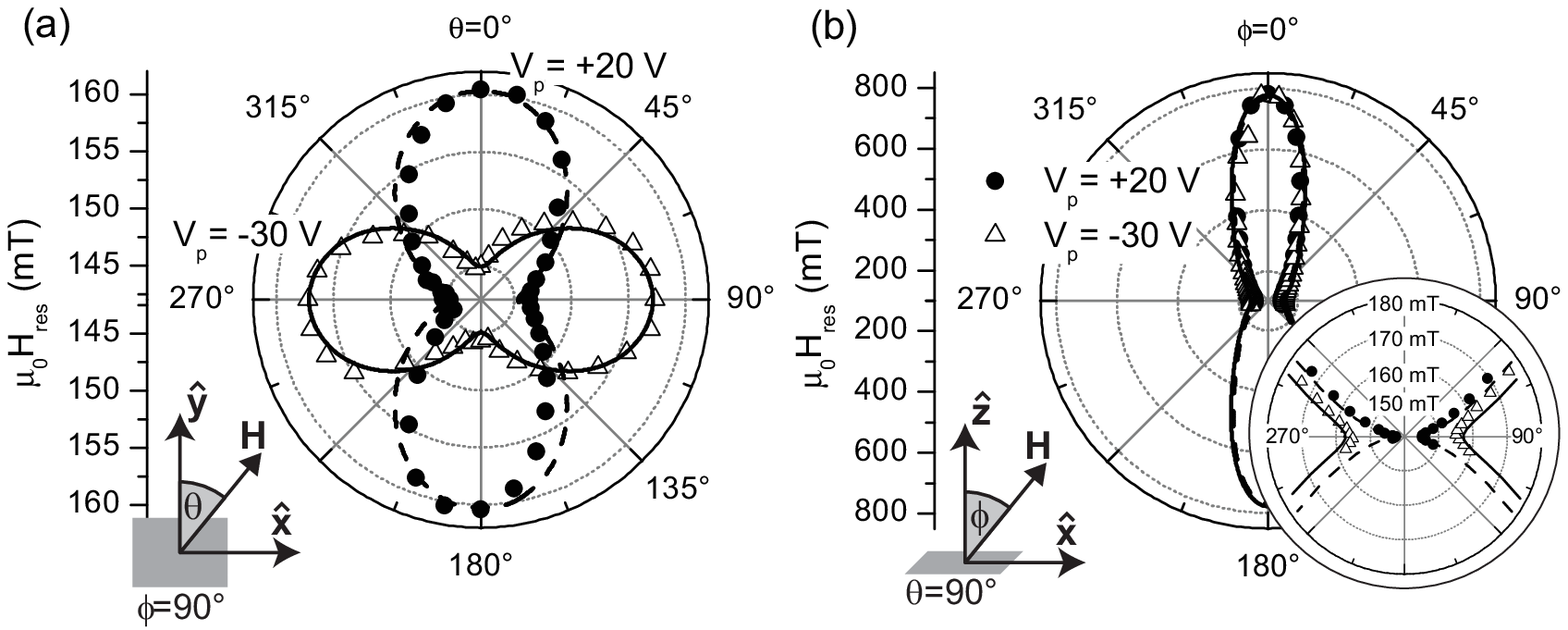}\\
  \caption{(a) The symbols show the FMR resonance field $\mu_0 H_\mathrm{res}(\theta)$ obtained for a constant actuator voltage $\Vp=\unit{-30}{\volt}$ (open triangles) and $\Vp=\unit{+20}{\volt}$ (solid circles) as a function of the orientation $\theta$ of $\mathbf{H}$ in the sample plane. A uniaxial (180\degree periodic) anisotropy of $\mu_0 H_\mathrm{res}(\theta)$ is observed for both voltages. However, the easy axis is rotated by 90\degree as $\Vp$ is changed from \unit{+20}{\volt} to \unit{-30}{\volt}. The full lines show the resonance fields simulated using the anisotropy constants from Eq.~\eqref{eq:constants}. (b) Corresponding experiments with $\mathbf{H}$ rotated in the \y plane (from within the Ni film plane to out-of-plane). Regardless of \Vp, a strong uniaxial anisotropy with the hard axis perpendicular to the film plane is observed. The inset shows that the resonance fields for $\mathbf{H}$ in the film plane ($\phi=90\degree$ and $\phi=270\degree$) are still shifted as a function of \Vp in accordance to the data shown in (a) for $\theta=90\degree$ and $\theta=270\degree$.}\label{fig:panel3}
\end{figure}
In this Section, we quantitatively determine the magnetic anisotropy of our \FPH structure using FMR measurements. Figure~\ref{fig:panel2}(a) shows four FMR spectra recorded at room temperature with \Hy and a voltage $\Vp \in \{\unit{-30}{\volt},\unit{0}{\volt},\unit{30}{\volt},\unit{90}{\volt}\}$ applied to the actuator, respectively. Each spectrum shows one strong ferromagnetic resonance which -- due to the magnetic field modulation and lock-in detection -- has a lineshape corresponding to the first derivative of a Lorentzian line~\cite{Poole1996}. The resonance field $\mu_0H_\mathrm{res}$ of a given FMR line is determined as the arithmetic mean of its maximum and minimum and depicted by the full squares in Fig.~\ref{fig:panel2}(a). The figure shows that, with the external magnetic field \Hy, $\mu_0H_\mathrm{res}$ is shifted to higher magnetic fields for increasing $\Vp$, while the lineshape is not significantly changed. According to Eqs.~\eqref{eq:Ftot} and~\eqref{eq:FMR_motion}, this implies that the \y direction becomes increasingly harder as $\Vp$ is increased.

To determine the evolution of $\mu_0 H_\mathrm{res}$ with \Vp in more detail, we recorded FMR spectra similar to those shown in Fig.~\ref{fig:panel2}(a) for \Vp increasing from $\unit{-30}{\volt}$ to $\unit{+90}{\volt}$ (upsweep) and decreasing back to $\unit{-30}{\volt}$ (downsweep) in steps of $\Delta\Vp=\unit{10}{\volt}$. For \Hy, we obtain the resonance fields shown by the squares in Fig.~\ref{fig:panel2}(b). Here, full squares depict the upsweep and open squares the downsweep of \Vp. As discussed in the context of Fig.~\ref{fig:panel2}(a), Fig.~\ref{fig:panel2}(b) shows that the FMR resonance field for \Hy increases with increasing \Vp and decreases with decreasing \Vp. The small hysteresis between up- and downsweep is due to the hysteretic expansion of the actuator~\cite{manual:piezo}. Carrying out the same series of FMR measurements with \Hx yields the resonance fields shown by the triangles in Fig.~\ref{fig:panel2}(b). For this magnetic field orientation, $\mu_0 H_\mathrm{res}$ decreases for increasing \Vp and vice versa. In terms of magnetic anisotropy we thus can conclude that the \x direction becomes easier for increasing \Vp while at the same time the \y direction simultaneously becomes harder.

For a more quantitative discussion we have also plotted the behavior expected from Eqs.~\eqref{eq:Ftot} and~\eqref{eq:FMR_motion} as solid lines in Fig.~\ref{fig:panel2}(b). These lines show the resonance fields obtained by assuming a linear, non-hysteretic voltage-strain relation [cf. Eq.~\eqref{eq:strain_voltage}] and solving Eq.~\eqref{eq:FMR_motion} with $F_\mathrm{tot}$ from Eq.~\eqref{eq:Ftot}. We use a saturation magnetization $M_\mathrm{s}=\unit{370}{\kilo\ampere\per\meter}$ as determined by SQUID measurements and a $g$-factor of 2.165~\cite{meyer:1961}.
Considering the fact that the actuator expansion saturates at high voltages and the data in Fig.~\ref{fig:panel2}(b) only show a minor loop of the full actuator swing, the simulation is in full agreement to the experimental results. This demonstrates that for $\Vp<\unit{0}{\volt}$ the \y-direction is magnetically easier than the \x-direction, while for $\Vp>\unit{0}{\volt}$ the \x-direction is easier than the \y-direction. Moreover, at $\Vp=\unit{0}{\volt}$, both measurement and simulation yield resonance fields of $\mu_0H_\mathrm{res}\approx\unit{150}{\milli\tesla}$ for \Hy as well as \Hx. Thus, at this voltage, both orientations \x and \y are equally easy in our sample.

To quantitatively determine the full magnetic anisotropy as a function of $\Vp$, we recorded FMR traces at constant \Vp for several different $\mathbf{H}$ orientations. The FMR resonance fields thus determined in experiment are shown in Fig.~\ref{fig:panel3} together with simulations according to Eqs.~\eqref{eq:Ftot} and~\eqref{eq:FMR_motion}. The open triangles in Fig.~\ref{fig:panel3} represent the resonance fields obtained for $\Vp=\unit{-30}{\volt}$ and the full circles those obtained for $\Vp=\unit{+20}{\volt}$. Note that in the polar plots in Fig.~\ref{fig:panel3}, the distance of $\mu_0 H_\mathrm{res}$ from the coordinate origin is an indicator of the magnetic hardness, with easy directions corresponding to small distances and hard directions to large distances.

If $\mathbf{H}$ is rotated in the film plane [cf. Fig.~\ref{fig:panel3}(a), $\phi=90\degree$], the obtained $\mu_0H_\mathrm{res}(\theta)$ shows minima at $\theta=0\degree$ and $\theta=180\degree$ for $\Vp=\unit{-30}{\volt}$ and at $\theta=90\degree$ and $\theta=270\degree$ for $\Vp=\unit{+20}{\volt}$, respectively. Thus, a clear 180\degree periodicity of the resonance fields and hence a uniaxial magnetic anisotropy is observed for both \Vp.  As the orientations $\theta$ corresponding to minima of $\mu_0H_\mathrm{res}$ for one voltage coincide with maxima for the other voltage, we conclude that the direction of the easy axis is rotated by 90\degree if \Vp is changed from $\Vp=\unit{-30}{\volt}$ to $\Vp=\unit{+20}{\volt}$. This is exactly the behaviour expected according to Figs.~\ref{fig:panel1}(c) and~\ref{fig:panel1}(e).

If $\mathbf{H}$ is rotated from within the film plane ($\phi=90\degree$, $\theta=90\degree$) to perpendicular to the film plane ($\phi=0\degree$, $\theta=90\degree$), we obtain the resonance fields shown in Fig.~\ref{fig:panel3}(b). In this case, we observe a strong uniaxial anisotropy with a hard axis perpendicular to the film plane regardless of \Vp, stemming from the \Vp-independent contribution $F_\mathrm{demag}$ to $F_\mathrm{tot}$. The inset in Fig.~\ref{fig:panel3}(b) shows that for $\mathbf{H}$ in the film plane, the resonance fields for $\Vp=\unit{-30}{\volt}$ and $\Vp=\unit{+20}{\volt}$ are shifted by approximately \unit{10}{\milli\tesla} in full accordance to the data in Fig.~\ref{fig:panel3}(a).

As discussed in Section~\ref{sec:theory}, the \Vp dependence of the resonance field is given by a \Vp dependence of $F_\mathrm{magel}$. We thus evaluate the measurement data shown in Figs.~\ref{fig:panel3}(a,b) using an iterative fitting procedure of $K_\mathrm{u,magel,y}$ in Eq.~\eqref{eq:Ftot} to fulfill Eq.~\eqref{eq:FMR_motion}. We obtain
\begin{eqnarray}\label{eq:constants}
  K_\mathrm{u,magel,y}(\unit{-30}{\volt})/M_\mathrm{s} &=& \unit{-3.4}{\milli\tesla} \\
  \nonumber K_\mathrm{u,magel,y}(\unit{+20}{\volt})/M_\mathrm{s} &=& \unit{4.4}{\milli\tesla}\;.
\end{eqnarray}
The resonance fields calculated using these anisotropy fields as well as Eqs.~\eqref{eq:FMR_motion} and~\eqref{eq:Ftot} are depicted by the lines in Fig.~\ref{fig:panel3}. The good agreement between simulation and experiment shows that $F_\mathrm{tot}$ given by Eq.~\eqref{eq:Ftot} is sufficient to describe the magnetic anisotropy of the Ni film.

In summary, the FMR experiments conclusively demonstrate that it is possible to invert the in-plane magnetic anisotropy of our hybrid structure, i.e. to invert the sign of $ K_\mathrm{u,magel,y}$ \textit{solely by changing \Vp.} As the FMR experiment furthermore allows to quantitatively determine all contributions to the free energy, Eq.~\eqref{eq:Ftot}, the orientation of the magnetization vector $\mathbf{M}$ in our sample can be calculated a priori for arbitrary $\mathbf{H}$ and \Vp. However, it is not possible to directly measure the magnetization orientation as a function of \Vp with FMR. To demonstrate that the piezo-voltage control of magnetic anisotropy indeed allows for a voltage control of $\mathbf{M}$, we now turn to magnetometry.

\section{Magnetometry}\label{sec:squid}
\begin{figure}
  \centering
  \includegraphics[]{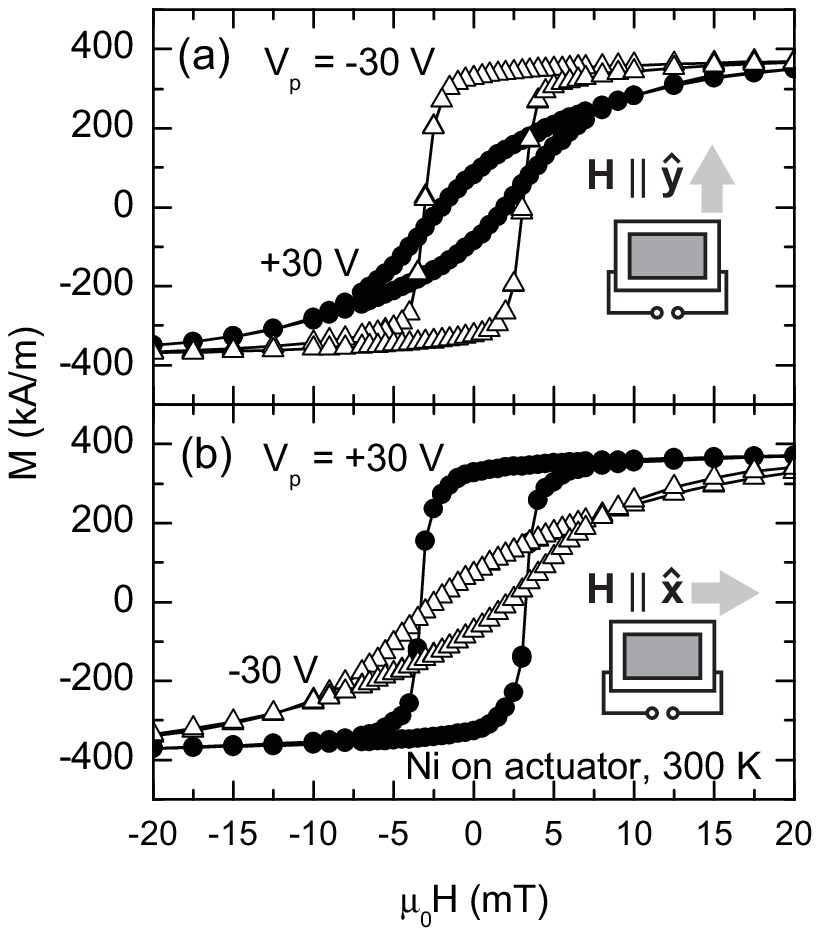}\\
  \caption{(a) $M(H)$-loops recorded with \Hy show a higher remanent magnetization for $\Vp=\unit{-30}{\volt}$ than for $\Vp=\unit{+30}{\volt}$, thus the \y-axis is magnetically easier for $\Vp=\unit{-30}{\volt}$ than for $\Vp=\unit{+30}{\volt}$. (b) The \x-axis is magnetically easier for $\Vp=\unit{+30}{\volt}$ than for $\Vp=\unit{-30}{\volt}$.}\label{fig:panel4}
\end{figure}
\begin{figure}
\centering
  \includegraphics[]{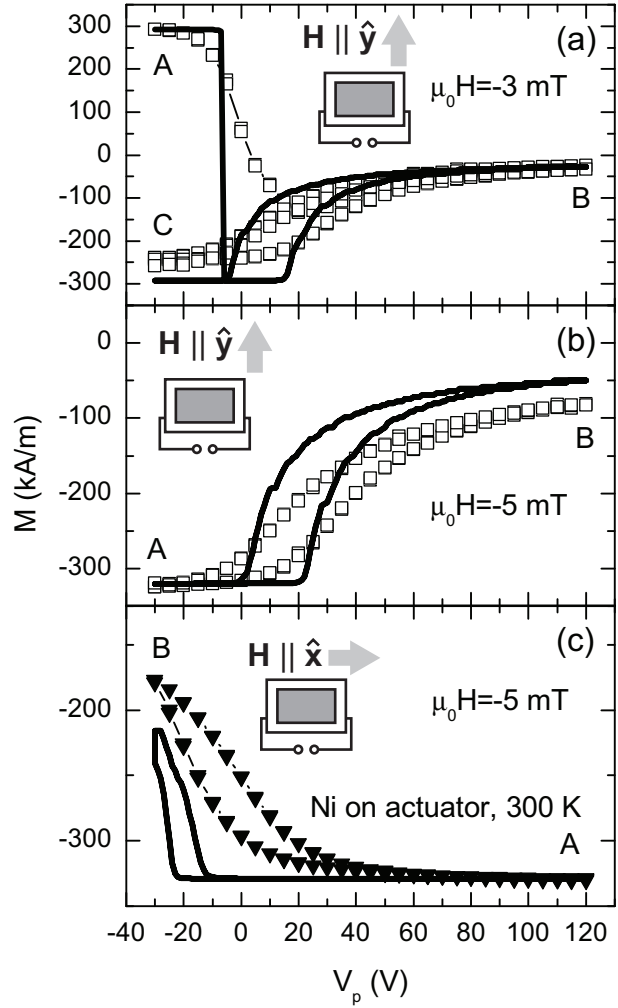}\\
  \caption{SQUID $M(\Vp)$-loops show the projection $M$ of the magnetization $\mathbf{M}$ onto the direction of the applied magnetic field $\mathbf{H}$ as a function of $\Vp$. The symbols represent the experimental data and the lines show the simulation of $M$ resulting from a minimization of $F_\mathrm{tot}$ [Eq.~\eqref{eq:Ftot}], with $\varepsilon_2$ determined using a strain gauge to explicitly take into account the actuator hysteresis. (a) For \Hy with $\mu_0 H=\unit{-3}{\milli\tesla}$, $\mathbf{M}$ exhibits an \textit{irreversible} rotation from A to B followed by a \textit{reversible} rotation from B to C. (b), (c) For \Hy and \Hx with $\mu_0 H=\unit{-5}{\milli\tesla}$, $\mathbf{M}$ exhibits a \textit{reversible} rotation from A to B and back to A.}\label{fig:panel5}
\end{figure}
\begin{figure}
\centering
  \includegraphics[]{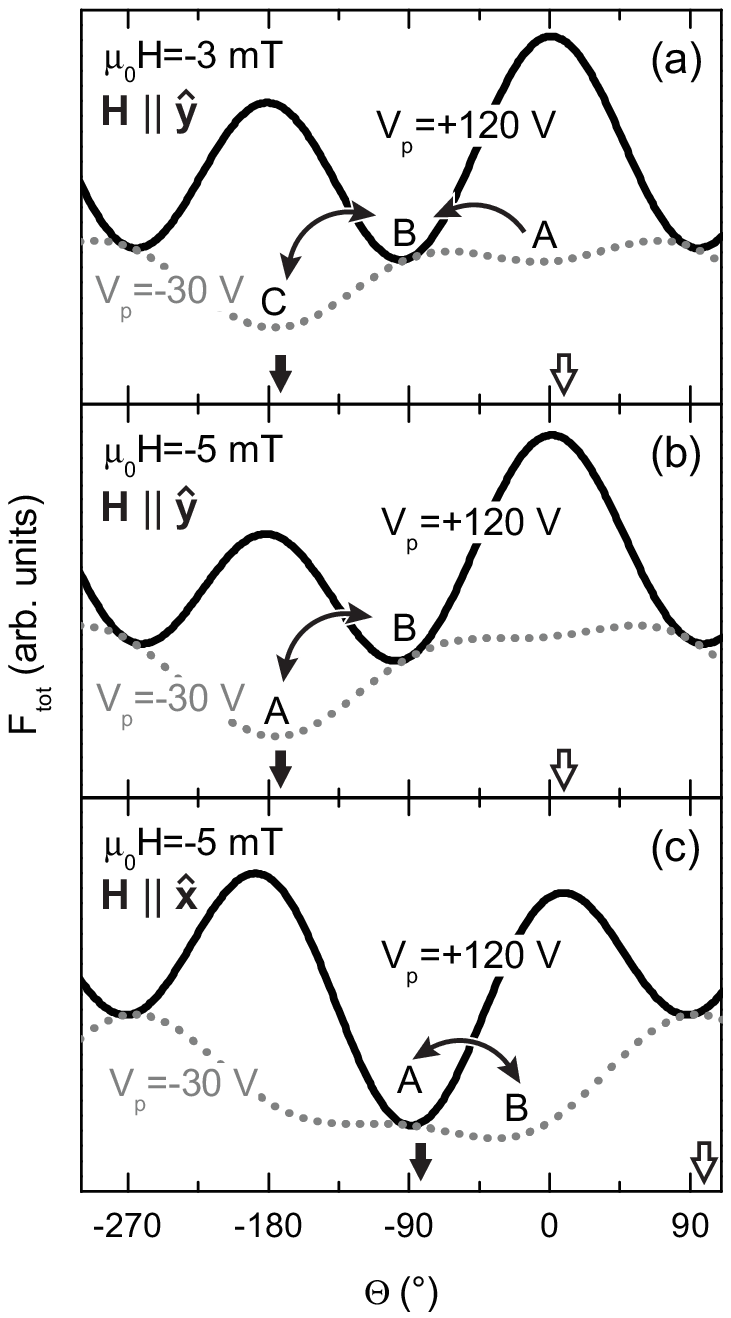}\\
  \caption{Calculated free energy contours in the film plane at the points of the $M(\Vp)$-loop depicted by capital letters in Fig.~\ref{fig:panel5} (solid lines: $\Vp=\unit{+120}{\volt}$, dotted lines: $\Vp=\unit{-30}{\volt}$). To clarify the lifting of the free energy degeneracy by the magnetic field, an angle of 10\degree between the magnetic field and the $\x$ or $\y$ axis was assumed. The open downward-oriented arrows depict the orientation of $\mathbf{H}$ during the field preparation at \unit{7}{\tesla} and the closed downward-oriented arrows depict the orientation of $\mathbf{H}$ during the actual measurement. Capital letters indicate the equilibrium $\mathbf{M}$ orientation at the corresponding positions in Fig.~\ref{fig:panel5}.  (a) Subsequent to the field preparation at \unit{-30}{\volt}, $\mathbf{M}$ resides in a \textit{local} minimum of $F_\mathrm{tot}$  (point A) at $\mu_0H=\unit{-3}{\milli\tesla}$ and rotates to the \textit{global} minimum of $F_\mathrm{tot}$ at (point B) as \Vp is increased to $\Vp=\unit{+120}{\volt}$. Sweeping the voltage from \unit{+120}{\volt} to \unit{-30}{\volt} now results in $\mathbf{M}$ following the \textit{global} minimum of $F_\mathrm{tot}$. (b), (c)  For $\mu_0H=\unit{-5}{\milli\tesla}$, $\mathbf{M}$ follows the \textit{global} minimum of $F_\mathrm{tot}$ as \Vp is changed.}\label{fig:panel6}
\end{figure}
In this Section, we show that it is possible to not only change the magnetic anisotropy but to deliberately \textit{irreversibly} and/or \textit{reversibly} rotate $\mathbf{M}$ simply by changing \Vp. To this end, we need to employ an experimental technique that is directly sensitive to the orientation of $\mathbf{M}$ rather than to magnetic anisotropies. Here we used SQUID magnetometry to record the projection $m$ of the total magnetic moment $\mathbf{m}$ onto the direction of the external magnetic field $\mathbf{H}$.

In a first series of experiments we recorded $m$ as a function of the external magnetic field magnitude $\mu_0H$ at fixed orientations of $\mathbf{H}$ and fixed voltages \Vp at $T=\unit{300}{\kelvin}$. Figure~\ref{fig:panel4}(a) shows $M=m/V$ measured with \Hy as a function of the external magnetic field strength at constant voltage $\Vp=\unit{+30}{\volt}$ (full circles) and $\Vp={-30}{\volt}$ (open triangles). The Ni film shows a rectangular $M(H)$ loop for $\Vp=\unit{-30}{\volt}$, while for $\Vp=\unit{+30}{\volt}$ the remanent magnetization is lowered by a factor of approximately three. According to, e.g., Morrish~\cite{Morrish2001}, the rectangular loop for $\Vp=\unit{-30}{\volt}$ indicates a magnetically easy axis, while the smooth, s-shaped loop for $\Vp=\unit{+30}{\volt}$ indicates a magnetically harder axis. Thus Fig.~\ref{fig:panel4}(a) shows that the \y-direction is magnetically \textit{easier} for $\Vp=\unit{-30}{\volt}$ and \textit{harder} for $\Vp=\unit{+30}{\volt}$ -- as expected from the FMR experiments. Changing the orientation of $\mathbf{H}$ to \Hx yields the $M(H)$-loops shown in Fig.~\ref{fig:panel4}(b). Following the same line of argument we can conclude that the \x-direction is magnetically \textit{easier} for $\Vp=\unit{+30}{\volt}$ and \textit{harder} for $\Vp=\unit{-30}{\volt}$. Altogether these results show that we observe an in-plane anisotropy, the easy axis of which is parallel to \y for $\Vp=\unit{-30}{\volt}$ and parallel to \x for $\Vp=\unit{+30}{\volt}$. These observations are fully consistent with the FMR results, and corroborate the simple model shown in Fig.~\ref{fig:panel1}.

Before discussing further experimental magnetometry results, we note that the free energy minima shown in Fig.~\ref{fig:panel1}(c,e) as well as the in-plane resonance fields in Fig.~\ref{fig:panel3}(a) are degenerate by 180\degree. This degeneracy may lead to demagnetization due to domain formation if the polarity of $\Vp$ is repeatedly inverted as there are always two energetically equally favorable but opposite directions of $\mathbf{M}$. Thus, to achieve a reversible control of $\mathbf{M}$ orientation, the degeneracy of $F_\mathrm{tot}$ needs to be lifted. This can be achieved if a unidirectional anisotropy is superimposed on the uniaxial anisotropy. Regarding $F_\mathrm{tot}$ in Eq.~\eqref{eq:Ftot}, only $F_\mathrm{stat}$ possesses the desirable unidirectional anisotropy. Hence, a small but finite external magnetic field $\mathbf{H}\neq0$ can be used to lift the 180\degree degeneracy. This approach works for all $\mathbf{H}$ orientations, except for $\mathbf{H}$ exactly parallel \x or \y. For \Hx (with $H>0$), the easy axis parallel to \x in Fig.~\ref{fig:panel1}(c) is replaced by an easier positive \x direction and a harder negative \x direction, thus the degeneracy is lifted for $\Vp>0$. However, for $\Vp<0$ [cf. Fig.~\ref{fig:panel1}(e)] a magnetic field \Hx is orthogonal to the easy axis which thus remains degenerate. The same consideration holds for \Hy where we expect degenerate free energy minima for $\Vp>0$ and a preferred $\mathbf{M}$ orientation for $\Vp<0$. However, in experiment it is essentially impossible to orient $\mathbf{H}$ \textit{exactly} along \x or \y; any small misorientation between $\mathbf{H}$ and \x or \y is sufficient to lift the degeneracy.

We now turn to the experimental measurement of the voltage control of $\mathbf{M}$ orientation. To show that $\mathbf{M}$ can be rotated by varying $\Vp$ alone, we change $\Vp$ at constant external magnetic bias field $\mathbf{H}$. As SQUID magnetometry is limited to recording the projection of $\mathbf{M}$ on the direction of the external magnetic field $\mathbf{H}$, it is important to choose appropriate $\mathbf{H}$ orientations in the experiments. As evident from Figs.~\ref{fig:panel1} and \ref{fig:panel4}, \Hx and \Hy are the most interesting orientations of the external magnetic field. In view of the circumstance discussed in the previous paragraph, we applied $\mathbf{H}$ to within 1\degree of \x or \y, respectively. We still refer to these orientations of $\mathbf{H}$ as \Hx and \Hy for simplicity, but take into account a misalignment of 1\degree in the calculations. As the experimental results and the corresponding simulations will show, this slight misalignment is sufficient to lift the degeneracy in the free energy regardless of $\Vp$.

Recording $M$ as a function of $\Vp$ for two complete voltage cycles $\unit{-30}{\volt}\leq \Vp \leq \unit{+120}{\volt}$ with \Hy and $\mu_0H=\unit{-3}{\milli\tesla}$ yields the data points shown in Fig.~\ref{fig:panel5}(a). Since \Hy, $M$ is the projection of $\mathbf{M}$ to the \y direction in the experiment. Prior to the first voltage sweep starting at point A, the voltage was set to $\Vp=\unit{-30}{\volt}$ and the sample was magnetized to a single domain state by applying $\mu_0 H=\unit{+7}{\tesla}$ along \y which is the easy axis for $\Vp=\unit{-30}{\volt}$. The magnetic field was then swept to $\mu_0 H=\unit{-3}{\milli\tesla}$ which is close to but still below the coercive field of the rectangular loops [cf. Fig.~\ref{fig:panel4}] and the acquisition of the $M(\Vp)$ data was started. The fact that $M$ is positive at first (starting from point A), while the field is applied along the negative \y direction, shows that $\mathbf{M}$ and $\mathbf{H}$ are essentially antiparallel at first. Upon increasing \Vp in steps of $\unit{+5}{\volt}$ to $\Vp=\unit{+120}{\volt}$ (point B), $M$ vanishes, which indicates an orthogonal orientation of $\mathbf{M}$ and $\mathbf{H}$. Upon reducing $\Vp$ to its initial value of \unit{-30}{\volt}, $M$ becomes negative (point C), evidencing a parallel orientation of $\mathbf{M}$ and $\mathbf{H}$. The \Vp cycle was then repeated once more, with $M$ now reversibly varying between its negative value at point C and zero at B.  Hence the evolution of $M$ is qualitatively different in the first and in the second \Vp cycle, with an \textit{irreversible} $\mathbf{M}$ rotation in the first cycle and a \textit{reversible} $\mathbf{M}$ rotation in the second cycle. This behaviour is expected from the magnetic free energy [cf. Eq.~\eqref{eq:Ftot}] as the preparation (point A) yields $\mathbf{M}$ in a metastable state ($\mathbf{M}$ antiparallel $\mathbf{H}$).

After a renewed preparation with $\mu_0 H=\unit{+7}{\tesla}$ at $\Vp=\unit{-30}{\volt}$, we repeated the experiment with the external magnetic field \Hy but with a slightly larger magnitude $\mu_0 H=\unit{-5}{\milli\tesla}$ and again recorded $M$ for a complete voltage cycle. The magnetic field magnitude of $\mu_0 H=\unit{-5}{\milli\tesla}$ was chosen as it exceeds the coercive field (cf. Fig.~\ref{fig:panel4}) while keeping the influence of the Zeemann term in Eq.~\eqref{eq:Ftot} on the total magnetic anisotropy comparable to the influence of the strain-induced anisotropies [cf. Eq.~\eqref{eq:constants}]. The experimental data are shown by the symbols in Fig.~\ref{fig:panel5}(b). Here a parallel orientation of $\mathbf{M}$ and $\mathbf{H}$ is already observed at point A ($\Vp=\unit{-30}{\volt}$) and $\mathbf{M}$ rotates reversibly by approximately 70\degree towards point B ($\Vp=\unit{+120}{\volt}$) and back when \Vp is reduced to $\unit{-30}{\volt}$ again (point A).

To complete the picture, we repeated the $M(\Vp)$ experiment again, but now applied \Hx. The magnetic preparation with $\mu_0H=\unit{7}{\tesla}$ was now performed at $\Vp=\unit{+120}{\volt}$ to ensure that the preparation field was again applied along an easy axis. The data  recorded subsequently at $\mu_0H=\unit{-5}{\milli\tesla}$  [cf. Fig.~\ref{fig:panel5}(c)] show a reversible rotation of $\mathbf{M}$ by approximately 60\degree between points A, B ($\Vp=\unit{-30}{\volt}$) and back to A.

To quantitatively simulate the evolution of $M$ with \Vp depicted in Fig.~\ref{fig:panel5}, we again took advantage of the fact that the free energy of the Ni film is known from the FMR experiments. To account for the hysteresis of the actuator expansion which is responsible for hysteresis in the $M(\Vp)$ loops in Fig.~\ref{fig:panel5}, we hereby used $\varepsilon_2$ as measured using a Vishay general purpose strain gauge in the voltage range of $\unit{-30}{\volt}\leq\Vp\leq\unit{+120}{\volt}$. The $\varepsilon_2$ data thus obtained are in accordance with the actuator data sheet.
Using $\varepsilon_2(\Vp)$ as well as the material constants given above we obtained in-plane free energy contours for each voltage \Vp. Figure~\ref{fig:panel6} exemplarily shows such free energy contours at the voltages depicted by the capital letters in Fig.~\ref{fig:panel5}. To clearly visualize the effect of $\mathbf{H}$ misalignment with respect to the sample coordinate system on $F_\mathrm{tot}$, the plots in Fig.~\ref{fig:panel6} were calculated assuming a misalignment in the in-plane orientation of the external magnetic field $\theta$ of 10\degree, so that $\theta=10\degree$  for \Hy and $\theta=100\degree$ for \Hx. Figure~\ref{fig:panel6} clearly shows that under these conditions the local minima of $F_\mathrm{tot}$ are non-degenerate for $\Vp=\unit{-30}{\volt}$ and $\Vp=\unit{+120}{\volt}$.

To determine the orientation of $\mathbf{M}$, we traced the minimum of the total free energy $F_\mathrm{tot}$ as a function of \Vp. This was done by minimizing Eq.~\eqref{eq:Ftot} with respect to the in-plane orientation $\Theta$ of $\mathbf{M}$, whilst assuming that $\Phi=90\degree$ due to the strong shape anisotropy.

To simulate the experimental $M(\Vp)$-loops [cf. Fig.~\ref{fig:panel5}], we assume a more realistic misalignment of 1\degree in $\theta$ and numerically minimize $F_\mathrm{tot}(\Theta)$ as a function of $\Vp$. For this, we set the initial value of $\Theta$ antiparallel to the external magnetic field for $\mu_0H=\unit{-3}{\milli\tesla}$ and parallel to the external magnetic field for $\mu_0H=\unit{-5}{\milli\tesla}$. Minimizing $F_\mathrm{tot}(\Theta)$ determines the $\mathbf{M}$ orientation which we project onto the \y or \x axis to yield $M$. For this, the magnitude of $\mathbf{M}$ is chosen to give a good fit to the experimental data at points A and is assumed to remain constant independent of \Vp.

The resulting simulations of $M$ are shown by the solid lines in Fig.~\ref{fig:panel5}. The simulation yields the experimentally observed $\textit{irreversible}$ $\mathbf{M}$ rotation during the first voltage sweep in Fig.~\ref{fig:panel5}(a) as well as the $\textit{reversible}$ $\mathbf{M}$ rotations in the second voltage sweep in Fig.~\ref{fig:panel5}(a) and in Figs.~\ref{fig:panel5}(b,c). The simulated total swing of $M$ is in excellent agreement with the experimental results for \Hy and in good agreement for \Hx. The fact that the experimental results exhibit a more rounded shape than the simulation is attributed to domain formation during the magnetization reorientation, which is neglected in the simulation.

Taken together, the free energy density of Eq.~\eqref{eq:Ftot} with the contributions quantitatively determined from FMR and our simple simulation of $M(\Vp)$ yield excellent agreement with experiment. In particular, we would like to emphasize that for the experiment and the simulation the application of \Vp leads to a \textit{rotation} of $\mathbf{M}$ and not to a decay into domains. This is evident from the fact that in Figs.~\ref{fig:panel5}(a) and \ref{fig:panel5}(b), a large projection of $\mathbf{M}$ onto \y is accompanied by a small projection onto \x and vice versa. Combining FMR and SQUID magnetometry, we thus have unambiguously and quantitatively demonstrated that $\mathbf{M}$ can be rotated reversibly by about 70\degree at room temperature solely via the application of appropriate voltages \Vp.

\section{Summary}
The FMR measurements show that the in-plane anisotropy of Ni/piezoactor hybrids can be inverted if the polarity of the voltage applied to the actuator is changed.  The magnetometry results corroborate this result and furthermore show that it is possible to irreversibly or reversibly rotate $\mathbf{M}$ solely by changing \Vp. For an \textit{irreversible} $M(\Vp)$ rotation, an appropriate preparation of $\mathbf{M}$ using a magnetic field sweep is necessary -- i.e. $\mathbf{M}$ must be aligned in a local free energy minimum. It then can be rotated out of this minimum by changing \Vp and the corresponding $\mathbf{M}$ orientation change can amount to up to 180\degree.
However, this voltage control of $\mathbf{M}$ is irreversible in the sense that $\mathbf{M}$ can not be brought back into the original orientation by changing \Vp alone. Rather, a second appropriate magnetic field sweep is required to align $\mathbf{M}$ into its original orientation.
In contrast, the \textit{reversible} $M(\Vp)$ reorientation requires a preparation of $\mathbf{M}$ only once. In this case, $\mathbf{M}$ is oriented along a global free energy minimum, and can be rotated within up to 70\degree (90\degree in the ideal case) at will by applying an appropriate \Vp. The $M(\Vp)$-loops simulated by minimizing the single-domain free energy [Eq.~\eqref{eq:Ftot}] are in excellent agreement with experiment, showing that the $\mathbf{M}$ orientation at vanishing magnetic field strengths can still be accurately calculated from the free energy determined from FMR. Finally, we would like to point out that the hysteretic expansion/contraction of the actuator visible as a hysteresis in $M(\Vp)$ [cf. Fig.~\ref{fig:panel5}] in particular also leads to distinctly different $M(\Vp=0)$, depending on the \Vp history. Thus, our data also demonstrate that a remanent $\mathbf{M}$ control is possible in our ferromagnetic thin film/piezoelectric actuator hybrids.

\ack
The work at the Walter Schottky Institut was supported by the Deutsche Forschungsgemeinschaft (DFG) via SFB 631. The work at the Walther-Meissner-Institut was supported by the DFG via SPP 1157 (Project No.~GR 1132/13), DFG Project No.~GO 944/3-1 and the German Excellence Initiative via the "Nanosystems Initiative Munich (NIM)".

\section*{References}
\bibliographystyle{prsty2}
\bibliography{literature}
\end{document}